\documentclass[a4paper,12pt]{article}
\usepackage{amsmath}
\usepackage[bf]{caption}
\usepackage{indentfirst}
\usepackage{graphicx}

\newcommand{\kk}{$\rm K \overline{K}$}
\newcommand{\kbk}{$\rm K^0 \, \overline{K} \, \! ^0$}
\newcommand{\ph}{$\pi\eta$}
\newcommand{\hpp}{$\eta\pi^+$}
\newcommand{\hp}{$\eta\pi$}
\newcommand{\hpeq}{\eta\pi}
\newcommand{\pipi}{$\pi\pi$}
\newcommand{\kkeq}{\rm K \overline{K}}
\newcommand{\pheq}{\pi\eta}
\newcommand{\pp}{$\rm p \overline{p}$}
\newcommand{\qq}{$q \overline{q}$}
\newcommand{\qqqq}{$q \overline{q} q \overline{q}$}
\newcommand{\az}{$a_0$}
\newcommand{\am}{$a_0(980)$}
\newcommand{\fz}{$f_0$}
\newcommand{\fm}{$f_0(980)$}
\newcommand{\ad}{$a_0(1450)$}
\newcommand{\ameq}{a_0(980)}
\newcommand{\fmeq}{f_0(980)}

\newcommand{\pmp}{$-+$}
\newcommand{\pmm}{$--$}
\newcommand{\pmpeq}{-+}
\newcommand{\pmmeq}{--}
\begin{document}

\title{\bf Coupled channel study of \az{} resonances}
\author{\bf Agnieszka Furman and Leonard Le\'{s}niak \\*[5pt]
\small {\it{Henryk Niewodnicza\'nski Institute of Nuclear Physics,}}\\*[-5pt]
\small {\it{PL 31-342 Krak\'ow, Poland}}}

\maketitle
\vspace{-7.5cm}
\begin{tabbing}
put this on the right hand corner using tabbing so it looks\= \kill
\>{IFJ-1898/PH/2002}
\end{tabbing}
\vspace{6.5cm}

\begin{abstract}

The coupled channel model of the \am{} and \ad{} resonances has been 
constructed using the separable \ph{} and \kk{} interactions. 
We have shown that two S--matrix poles corresponding to the \am{} 
meson have significantly different widths in the complex energy 
plane.
The \kk{} to \ph{} branching ratio, predicted in our model near the 
\ad{} mass, is in agreement with the result of the Crystal Barrel 
Collaboration. 
The \kk{} interaction in the S--wave isovector state is {\it not} 
sufficiently attractive to create a bound \am{} meson. 
\end{abstract}

\maketitle 

\section{Introduction}

Properties of scalar mesons are 
intensively studied in many theoretical and 
experimental articles [1--3]. 
The number of known isovector resonances in the scalar meson sector 
is smaller than the corresponding number of isoscalar resonances 
since only two isospin $I=1$ states \am{} and \ad{} are listed in the 
last edition of the Review of Particle Physics \cite{pdg00}. 
The \am{} state lies close to the \kk{} threshold which can strongly 
influence the resonance shape in the \ph{} main decay channel. 
By comparing results given in Refs. \cite{BNL99} and \cite{CB98} one 
can notice some differences in experimental determinations of the \am{} 
mass and particularly of its width. 
More important mass and width differences of the \ad{} state found 
in the \pp{} annihilation by the Crystal Barrel Collaboration 
\cite{CB98} and by the OBELIX Collaboration \cite{OB98} have been 
pointed out by Montanet \cite{montanet00}. 

Internal structure of scalar mesons is not yet well understood. 
Many conflicting view--points in which scalars are treated as \qq{} 
or \qqqq{} states, two--meson quasi--bound states and mixed states 
including glueballs coexist (see, for example, [8--13]).

The \az{} resonances decay mainly to \ph{} and \kk{} channels and the 
\ad{} decays also to $\pi \eta'(958)$ \cite{CB97}. 
Studies of the decay channels are potentially a rich source of 
information about a dynamics of the meson--meson interactions. 
Unfortunately the \ph{} and the \kk{} phase shifts in the 
scalar--isovector state have not been experimentally determined although 
in several production experiments partial wave analyses have been 
performed \cite{BNL99,GAMS99,WA10200}. 
On the other hand these phase shifts or the meson--meson scattering 
amplitudes have been calculated using theoretical models [9, 17--19]. 
We should remark, however, that an experimental determination of the 
\ph{} phase shifts is more difficult than a derivation of the 
$\pi \pi$ phase shifts. 
In the second case a strong dominance of one pion exchange has helped 
to obtain the $\pi \pi$ scalar--isoscalar phase shifts from the 
amplitudes of the $\pi \pi$ production on nucleons using pion beams 
\cite{kaminski97,kaminski01}. 

Closeness of the \am{} to the $\rm K^+K^-$ and \kbk{} 
thresholds and existence of other scalar meson \fm{} at a very 
similar mass make possible an isospin mixing of both resonant states. 
The physical consequences of this phenomenon have been recently 
presented by several authors [22--24].
Theoretical investigations have been stimulated by new experimental 
results from Novosibirsk and Frascati on the radiative $\phi$ decays 
into $\ameq \gamma$ and $\fmeq \gamma$ [25--27].
A good experimental insight into the \az{}--\fz{} mixing can be 
achieved when an experimental effective mass resolution is 
significantly better than the mass difference between the 
$\rm K^+K^-$ and \kbk{} thresholds (about 8 MeV). 
Such a good resolution is also needed to measure 
the width of the \am{} meson with a sufficient precision. 

In the present paper we use the separable potential model 
formulated in Ref. \cite{lesniak96}. 
This is an extension of a similar two--channel model developed in Ref. 
\cite{kaminski94} to describe the $\pi \pi$ and \kk{} S--wave $I=0$ 
channel in which the $f_0(500)$ meson has been found. 
The mass and width of the $f_0(500)$ meson have been obtained by 
finding an appropriate pole of the S--matrix in the complex energy 
plane. 
The $\pi \pi$ and \kk{} experimental phase shifts served as input to 
fix the model parameters in the scalar--isoscalar sector. 

A simple application of the same procedure to the 
S--wave $I=1$ channels is not yet possible since we do not know 
experimental values of the \ph{} and \kk{} phase shifts. 
We can, however, fix four model parameters to fit values of masses and 
widths of two known \az{} resonances and use some experimental 
information about the \kk{} to \ph{} branching ratio for the \am{} 
to determine the 
remaining free parameter of the minimal version of the coupled channel 
(\ph{} and \kk{}) model. 
This analytic and unitary model has only five independent parameters 
but using it one can calculate elastic and inelastic \ph{} and \kk{} 
S--wave isovector amplitudes, \az{} coupling constants to 
two channels and positions of 
different S--matrix poles in the complex energy plane including those 
poles which have not been postulated as the experimental input. 
One can also answer an interesting question whether the \kk{} forces 
are sufficiently strong to create a bound $I=1$ S--wave state. 
Despite of its simplicity the model has an attractive possibility to 
check, confront and even correct phenomenological results obtained by 
different experimental groups in studies of the \az{} production 
processes. 
\section{Theoretical model}

Let us briefly describe our theoretical model 
\cite{lesniak96}. 
We consider two channels \ph{} (label 1) and
\kk{} (label 2).
The separable potentials describe interactions between mesons:
\begin{equation} \label{eq:v}
\langle\,\mathbf{p}\,|\,V_{ij}\,|\,\mathbf{q}\,\rangle=
\lambda_{ij}\,g_i(p)\,g_j(q)\;, \qquad i,j=1,2\;,
\end{equation}
where $\lambda_{ij}=\lambda_{ji}$ are the real coupling constants and 
$g_i$ are the form factors in Yamaguchi's form:
\begin{equation} \label{eq:gi}
g_i(p)=\sqrt{\,\frac{2\pi}{m_i}}\,\frac{1}{p^2+\beta_i^2}\;.
\end{equation}
In Eq. \eqref{eq:gi} $m_i$ are the reduced masses, $\beta_i$ are the
range parameters and $p$ is the relative momentum. 
The matrix $T$ of the scattering amplitudes satisfies the coupled 
channel Lippmann--Schwinger equation: 
\begin{equation}  \label{eq:Lipp-Schw}
\langle\,\mathbf{p}\, |\,T\,|\,\mathbf{q}\,\rangle=
\langle\,\mathbf{p}\,|\,V\,|\,\mathbf{q}\,\rangle+
\int\frac{ d^3s }{ (2\pi)^3 }
\langle\,\mathbf{p}\,|\,V\,|\,\mathbf{s}\,\rangle
\langle\,\mathbf{s}\,|\,G\,|\,\mathbf{s}\,\rangle
\langle\,\mathbf{s}\,|\,T\,|\,\mathbf{q}\,\rangle\;,
\end{equation}
where $T$, $V$, $G$ are $2 \times 2$ matrices. 
The elements of the diagonal matrix $G$ of propagators are given by:
\begin{equation}
G_i(s)=\frac{1}{E-E_i(s)+i\epsilon}\,,\qquad \epsilon \to 0+\;,
\end{equation}
where $s$ is the relative momentum, $E$ is the total energy and $E_i$ 
are the channels energies.
The T--matrix elements satisfying Eq. \eqref{eq:Lipp-Schw} can be 
written as 
\begin{equation}
\langle\,\mathbf{p}\,|\,T_{ij}\,|\,\mathbf{q}\,\rangle= 
                      g_i(p)\,t_{ij}\,g_j(q)\,, \quad i,j=1,2\,,
\end{equation}
where the matrix $t$ is expressed below:
\begin{equation}
t=(1-\lambda \, I)^{-1}\,\lambda\;.
\end{equation}
In this equation $\lambda$ is the symmetric matrix of the coupling
constants and $I$ is the diagonal matrix of integrals:
\begin{equation}
I_{ii}=\int\frac{d^3s}{(2\,\pi)^3}\,g_i(s)\,G_i(s)\,g_i(s)\;.
\end{equation}
The determinant of $(1-\lambda\,I)$, called the Jost function $D(E)$, 
can be written as a function of two channel centre of mass momenta 
$k_1$ and $k_2$:
\begin{equation} \label{eq:jost}
D(k_1,k_2)=D_1(k_1)\,D_2(k_2)-F(k_1,k_2)\;,
\end{equation}
where
\begin{equation} \label{eq:jost_i}
D_i(k_i)=1-\lambda_{ii}\,I_{ii}(k_i)\;, \qquad i=1,2\;,
\end{equation}
and
\begin{equation} \label{eq:fn_F}
F(k_1,k_2)=\lambda_{12}^2\,I_{11}(k_1)\,I_{22}(k_2)\;.
\end{equation}
The Jost function can be used to express all the S--matrix elements
in the following way:
\begin{equation} \label{eq:S}
 \begin{split}
S_{11}=\frac{D(-k_1,k_2)}{D(k_1,k_2)}\;,& \qquad 
S_{22}=\frac{D(k_1,-k_2)}{D(k_1,k_2)}\;, \\
S_{12}^2=S_{11}S_{22}&-\frac{D(-k_1,-k_2)}{D(k_1,k_2)}\;.
 \end{split}
\end{equation}

The matrix of potentials has five parameters: two $\beta_i$ and 
three independent $\lambda_{ij}$ parameters, which must be determined 
from data. 
In a case when masses and widths of two \az{} resonances are 
known a calculation of four parameters is rather simple. 
The positions of resonances in the complex energy plane coincide with 
poles of the S--matrix elements and with zeroes of the Jost function
$D(k_1,k_2)$. 
Thus four model parameters can be found by solving two complex 
equations:
\begin{equation} \label{eq:res}
D(k_1^r,\,k_2^r)=0 \, \textrm{ and } D(k_1^R,\,k_2^R)=0\,,
\end{equation}
where $k_{i}^{r}$ and $k_{i}^{R}$ denote the complex momenta in a 
channel $i$ related to the complex energy positions of the \am{} and 
\ad{} resonances, respectively.

There exists an ambiguity in determination of the momenta 
since a given resonance energy $E_p=M-i\, \Gamma/2$ ($M$ being a 
resonance mass and $\Gamma$ its width) can correspond to more than one 
resonance position in the complex plane of two momenta $k_1$ and 
$k_2$. 
The energy conservation relation is quadratic in momenta, thus 
resonances can be situated on different sheets labelled by signs of 
imaginary parts $Im\,k_1$ and $Im\,k_2$. 
For example, the \am{} resonance can be located on sheets \pmp{} 
or \pmm{}, called sheets II or III, respectively. 
\section{Experimental input and model parameters}

In \cite{CB98} the \am{} resonance production amplitudes have been 
parametrised using the $K$--matrix form reducing to the modified 
version of the Flatt\'e formula \cite{flatte76}: 
\begin{equation}\label{flatte}
 F_i(m)=\cfrac{N g_i}{m_0^2-m^2-i(\rho_1g_1^2+\rho_2g_2^2)}\;,\quad i=1,2\;,
\end{equation}
where $m$ is the \ph{} or \kk{} effective mass, $\rho_i=2k_i/m$ and 
$N$ is a constant. 
Using the parameters fitted in \cite{CB98}: $m_0=(999 \pm 2)$ MeV, 
$g_1=(324 \pm 15)$ MeV, and the ratio of coupling constants squared
$r=g_2^2/g_1^2=1.03 \pm 0.14$ we have calculated the positions 
of two poles of $F_i(m)$ at 
\begin{equation} \label{cb:our_poles}
 \begin{split}
M_1=&(1005 \pm 3) \textrm{ MeV, } \Gamma_1=(49 \pm 7)  
    \textrm{ MeV on sheet II } (\pmpeq)\\ \textrm{and} \quad 
M_2=&(985 \pm 7) \textrm{ MeV, } \Gamma_2=(92 \pm 13)  
    \textrm{ MeV on sheet III } (\pmmeq).
 \end{split}
\end{equation}
Let us note that the first pole is located above the \kk{} threshold while the 
second pole lies below it and the widths are different by a factor of 
about two. 
Similar results with very different widths can be found 
for solution B 
obtained in \cite{CB94} by the Crystal Barrel Collaboration. 
It can be shown that in the modified Flatt\'e model the difference of 
two widths is approximately proportional to the \kk{} coupling constant 
squared
\begin{equation}
\Gamma_2-\Gamma_1 \approx g_2^2 \cdot \cfrac{4 Re q_2}{m_0^2}\;,
\end{equation}
where $q_2$ is the \kk{} complex momentum corresponding to the pole on 
sheet III. 
Thus the widths of two poles have to be different. 
Here we should point out that some numbers in \eqref{cb:our_poles} do 
not coincide with the pole positions written in 
\cite{CB98} (also the sheet numbers seem to be interchanged).

The original Flatt\'e parametrisation has been used by the E852 
Collaboration in the study of the reaction 
$\pi^- p \to \eta \pi^+ \pi^- n$ measured at Brookhaven at 18.3 GeV/c 
\cite{BNL99}. 
The E852 Group has performed a fit of the observed \hpp{} effective mass 
distribution and obtained the mass position $M_r =(1001.3 \pm 1.9)$ 
MeV, the ratio of coupling constants $R=0.91 \pm 0.10$ and the value of 
dimensionless coupling constant $g_{\eta \pi}=0.243 \pm 0.015$, 
corresponding to the \hp{} width $\Gamma _{\hpeq}=81$ MeV calculated at 
the $M_r$ position. 
After an effective mass correction for a finite experimental 
resolution this value was reduced to $70 \pm 5$ MeV \cite{BNL99} which 
is equivalent to a reduction of the \hp{} coupling constant to 
$g_{\hpeq}=0.210 \pm 0.015$. 
Using this value of $g_{\hpeq}$ and the above numbers of $M_r$ and $R$ 
we have found positions of two poles at 
\begin{equation} \label{bnl:our_poles}
 \begin{split}
 M=&(1006 \pm 3) \textrm{ MeV, } \Gamma=(51 \pm 5) 
   \textrm{ MeV on sheet II } (\pmpeq)\\ \textrm{and} \quad
 M=&(988 \pm 5) \textrm{ MeV, } \Gamma=(89 \pm 10)
   \textrm{ MeV on sheet III } (\pmmeq).
 \end{split}
\end{equation}
These positions are in very good agreement with those calculated 
by us for the Crystal Barrel Collaboration parameters in 
\eqref{cb:our_poles}. 
Let us remark that the parameters of the \am{} meson cited in Table 
II \cite{BNL99} in the line labelled by E852(C) do not correspond to 
the pole position of the production amplitude, so $M_r$ and $\Gamma$ 
placed in that line should not be interpreted as the \am{} resonance 
mass and its width. 

The presence of two distinct poles related to the \am{} meson means 
that in principle one has two options while fixing the model parameters 
using Eq. \eqref{eq:res}. 
At first we shall discuss the case when the \am{} pole is located on 
sheet II at $M_1-i\,\Gamma_1/2=(1005-24.5\,i)$ MeV. 
We expect that the \ad{} resonance lies on sheet III since it is 
located far from the \ph{} and \kk{} thresholds. 
We take its mass as $M=1474$ MeV and the width as $\Gamma=265$ MeV. 

In the simplest version of our two channel model we need further 
experimental constraints to fix the fifth model parameter. 
We choose the range variable $\beta_1$ in the \ph{} channel as a free 
parameter to fit the \kk{} to \ph{} branching ratio defined as 
\begin{equation}
 BR=\cfrac{\int_{m_2}^{m_{max}} \rho_2 \,|F_2(m)|^2 \,dm }
          {\int_{m_1}^{m_{max}} \rho_1 \,|F_1(m)|^2 \,dm }\;,
\end{equation}
where $m_1$ and $m_2$ are lower bounds of the effective mass and 
$m_{max}$ is its upper bound. 
Using the K--matrix model of Ref. \cite{CB98} this ratio can be 
expressed as the ratio of the integrals over the transition cross section 
and over the \ph{} elastic cross section: 
\begin{equation}
BR=\cfrac
    {\int_{m_2}^{m_{max}}\sigma_{\pheq \to \kkeq}\,(m)\,k_1\,m\,dm}
    {\int_{m_1}^{m_{max}}\sigma_{\pheq}^{\,el}\,(m)\,k_1\,m\,dm}\;.
\end{equation} 

\begin{figure}[!ht]
\begin{center}
\includegraphics*[width=0.8\textwidth]{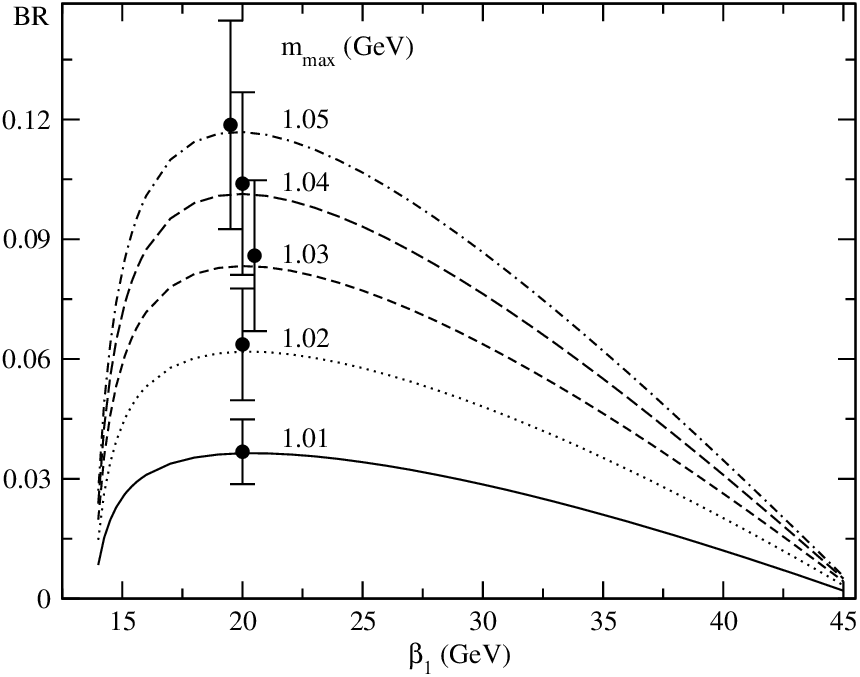}
\end{center}
\caption{Model dependence of the \kk{} to \ph{} branching ratio on the 
         range parameter $\beta_1$. 
         Points with errors are calculated BR values using the modified 
         Flatt\'e formula for the parameters of Ref. \cite{CB98}. 
	 The upper effective mass integration limit is denoted by 
	 $m_{max}$. }
\label{fig:br}
\end{figure}

The branching ratio calculated for the \am{} resonance on sheet II is 
shown in Fig. \ref{fig:br}. 
Here $m_1=m_{\pi}+m_{\eta}$ and $m_2=2 m_K$, where $m_{\pi}$,  
$m_{\eta}$ and $m_K$ are the pion, eta and kaon masses, respectively.
We see that BR depends strongly on the upper bound $m_{max}$. 
In order to avoid a strong interference of the 
      wide $a_0(1450)$ with the $a_0(980)$ resonance, the limits of $m_{max}$, 
      shown in Fig. 1, do not exceed $1.05$~GeV. 
      This interference is particularly important in 
      the $\rm K \overline{K}$ channel so the integration over the 
      $\rm K \overline{K}$ invariant masses higher than $1.05$~GeV 
      would lead to a significant distortion of the $a_0(980)$ branching 
      ratio.     

The variation of the branching ratio with $\beta_1$ is correlated with 
the dependence of the interchannel coupling constant $\lambda_{12}$ on 
$\beta_1$. 
Maximum of $\lambda_{12}^2$ lies for $\beta_1$ near $20$ GeV. 

The Particle Data Group \cite{pdg00} has calculated an average 
branching ratio from the results obtained by the Crystal Barrel 
Collaboration\cite{CB98} ($BR=0.23 \pm 0.05$) and by the WA102 Group 
\cite{WA10298} ($BR=0.166 \pm 0.01 \pm 0.02$). 
However, this procedure has to be taken with a great care since the 
value of BR depends very much on the integral limits. 
For the \pp{} annihilation at rest into $\pi^0 \pi^0 \eta$ the 
available \ph{} effective mass varies between 682 MeV at the 
$\pi^0 \eta$ threshold and the maximum value of about 1741 MeV. 
The WA102 Collaboration has studied the reaction $pp \to p f_1(1285)p$ 
with a subsequent decay of $f_1(1285)$ into the $\pi \pi \eta$ system. 
Then the maximum of the \ph{} effective mass equals to about 1147 MeV. 
We have recalculated the Crystal Barrel result 
$0.23 \pm 0.05$, corresponding to $m_{max}=1741$ MeV, by lowering the 
upper limit to $m_{max}=1147$ MeV. 
Then the number $BR=0.19 \pm 0.04$ was quite close to the WA102 
result cited above. 
Encouraged by this agreement we have made calculations restricting 
$m_{max}$ limit to values closer to the \kk{} threshold where the \am{} 
resonance dominates.
The results are shown in Fig. \ref{fig:br} by points with errors 
corresponding to the Crystal Barrel error 
$(0.05/0.23)\cdot 100 \% = 22 \% 
$.
A comparison with the theoretical curves following from our model 
indicates that the $\beta_1$ value should be chosen close to 20 GeV. 
Then the complete set of the potential parameters corresponding to 
the \am{} lying on sheet II and the \ad{} located on sheet III is 
following: 
\begin{equation} \label{eq:par}
\begin{split}
&\Lambda_{11}=2 \beta_1^3 \lambda_{11}=-0.032321, \\
&\beta_1=20.0 \textrm{ GeV,} \\
&\Lambda_{22}=2 \beta_2^3 \lambda_{22}=-0.068173, \\
&\beta_2=21.831 \textrm{ GeV}\qquad \textrm{and} \\
&\Lambda_{12}=
   2 (\beta_1 \beta_2 )^{3/2} \lambda_{12}=5.0152 \cdot 10^{-4}.
\end{split}
\end{equation}
\section{Predictions}
\subsection{\kk{}/\ph{} branching ratio}

Having established all the model parameters we can predict the 
\kk{}/\ph{} branching ratio in the \ad{} mass range. 
One can take equal limits $m_1=m_2$ and the \ad{} width $\Gamma=265$ 
MeV \cite{pdg00}. 
For a typical range of $m$ between
$m_1\approx 1474\textrm{ MeV }-\Gamma/2\approx 1340\textrm{ MeV}$ 
and 
$m_{max}\approx 1474\textrm{ MeV }+\Gamma/2\approx 1605\textrm{ MeV}$ 
the branching ratio equals to $0.98$. 
If we take wider limits like $m_1=1300$ MeV and $m_{max}=1741$ MeV, 
then the corresponding value is $0.78$. 
The Crystal Barrel Collaboration result \cite{CB98} $BR=0.88 \pm 0.23$ 
agrees quite well with both theoretical numbers following from our 
model. 
Thus there is a mutual consistency of two branching ratios calculated 
within our model for the \am{} and \ad{} mesons and the experimental 
numbers obtained by the Crystal Barrel Collaboration. 

Let us discuss for a moment the case when the \am{} pole is fixed on 
sheet III at $M_2-i\,\Gamma_2/2=(985-46i)$ MeV 
(see Eq. \eqref{cb:our_poles}). 
If we calculate the branching ratio BR as in Fig. \ref{fig:br} then the 
theoretical curves lie substantially lower (by a factor at least 2.5) 
than the recalculated experimental values. 
Thus in the following text this case will not be considered. 

We have also studied a dependence of the \am{} branching ratio on the 
input values of the \ad{} mass and width. 
Let us recall that the OBELIX Collaboration has observed the 
$a_0(1300)$ resonance at $M=1290$ MeV and $\Gamma=80$ MeV \cite{OB98}
\footnote{The OBELIX Collaboration has determined the \am{} branching 
ratio as $0.26 \pm 0.06$ \cite{OB00} in agreement with the Crystal 
Barrel Collaboration ratio $0.23 \pm 0.05$.}. 
If in our model we take the above values of $M$ and $\Gamma$ instead of 
$M=1474$ MeV and $\Gamma=265$ MeV then the resulting \am{} branching 
ratio is too small by a factor at least 3 for both positions of \am{} 
in \eqref{cb:our_poles} when it is compared with numbers represented by 
experimental points in Fig. \ref{fig:br}. 
Even if the \am{} mass would be lowered to 975 MeV as given in 
\cite{OB98} then the \am{} branching ratio should be even smaller. 
Hence we have found that, using our two channel model, it is difficult 
to link the parameters of the \am{} and the $a_0(1300)$ resonances 
obtained by the OBELIX Collaboration. 
\subsection{Poles and zeroes}

The relations \eqref{eq:S} between the Jost function 
and the S--matrix elements are very useful in analysis of the 
positions of the S--matrix poles and zeroes in the complex planes 
of momenta $k_1$ and $k_2$. 
A configuration of the poles and zeroes nearest to the physical region 
is important to understand a behaviour of the scattering amplitudes
\begin{equation} \label{eq:Tij}
T_{ij}=\cfrac{S_{ij}-\delta_{ij}}{2 i}\;,\quad i,j=1,2.
\end{equation}%

The positions of the S--matrix poles corresponding 
to our choice \eqref{eq:par} of model 
parameters are given in Table \ref{tab:poles} 
\begin{table}[ht]
\caption{Positions of the S--matrix poles (in units of MeV)}
\vspace{6pt}
\centering
\begin{tabular}{|r|r|r|r|r|r|r|} \hline
pole & $Re E$ & $Im E$  &$Rek_1$&$Imk_1$ &$Rek_2$&$Imk_2$\\ \hline
1    & 1005.0 &  -24.5  & 336.6 & -16.9  & -94.5 &  65.2 \\
2    &  991.5 &  -33.6  & 327.4 & -23.3  &  85.4 & -97.5 \\ \hline
3    & 1474.0 & -132.5  & 628.2 & -76.4  & 546.9 & -89.3 \\
4    & 1467.3 &   -6.9  & 623.4 &   4.0  &-539.1 &  -4.7 \\ \hline
\end{tabular} \label{tab:poles}
\end{table}%
\begin{figure}[!ht]%
\begin{center}
\includegraphics*[width=\textwidth]{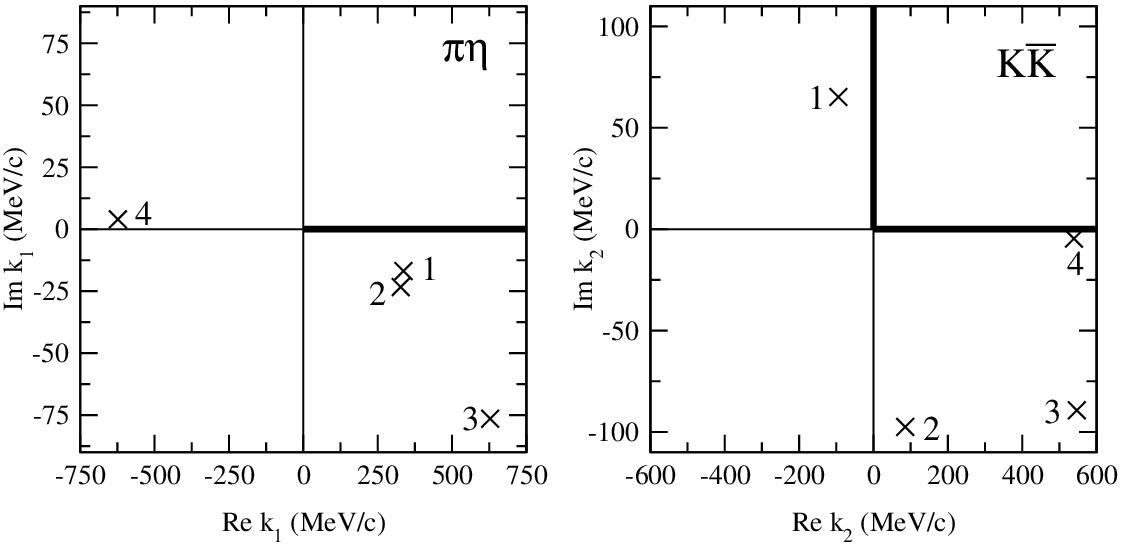}
\end{center}
\caption{Configuration of the S--matrix poles (denoted by crosses)
         in the \ph{} and \kk{} channels. 
         Poles 1 and 2 correspond to \am{} and poles 3 and 4 
         correspond to \ad{}. 
         The solid lines indicate the physical region.}
\label{fig:poles}
\end{figure}%
and graphically 
presented in Fig. \ref{fig:poles}. 
One can see that in addition to two postulated poles (1 and 3) 
the other two poles (2 and 4) are present. 
Near the \kk{} threshold two poles (1 and 2) play an important 
role but the additional pole 4, related to \ad{}, lies far 
from the physical region. 
It does not mean, however, that this pole has no 
influence on the scattering amplitudes near 1450 MeV. 
In fact, every zero of the Jost function has a twin zero which 
lies symmetrically with respect to the imaginary momentum axis 
in both planes. 
This fact results from the general relation 
$D(k_1,k_2)=D^*(-k_1^*,-k_2^*)$ satisfied by the Jost function. 
Thus a zero of the $S_{22}$ element%
\footnote{actually equivalent to a zero of $D(k_1,-k_2)$}
corresponding to a twin pole related to the pole 4 is localised very 
close and slightly above the physical axis in the \kk{} plane. 
Similarly in the \ph{} complex momentum plane a zero of the $S_{11}$ 
corresponding to the pole 4 lies slightly below the real axis 
and has an important influence on the \ph{} scattering amplitude. 
\subsection{Scattering amplitudes}

\begin{figure}[!ht]
\begin{center}
\includegraphics*[width=\textwidth]{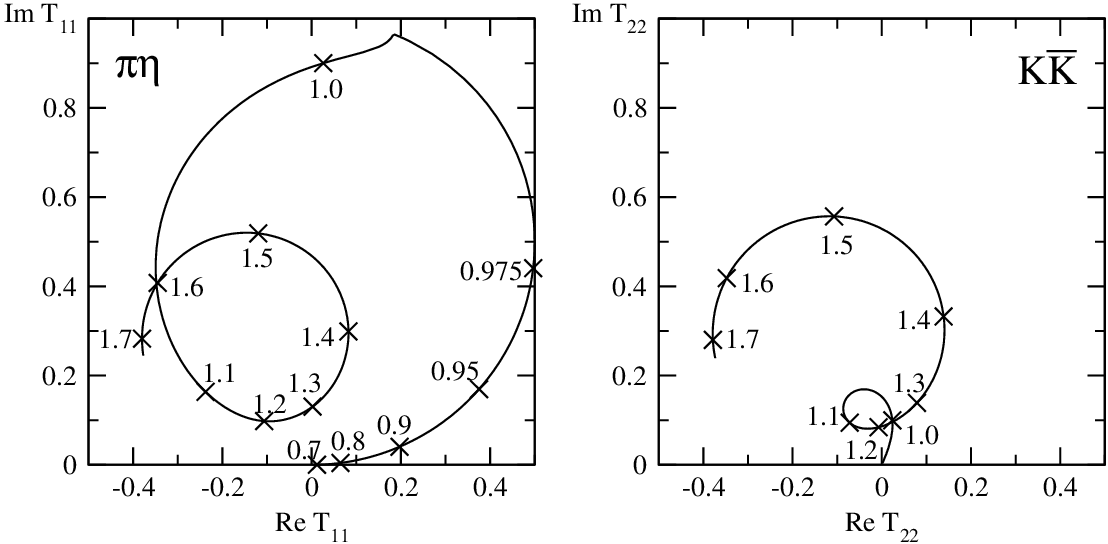}
\end{center}
\caption{Argand diagrams for the \ph{} and \kk{} elastic scattering 
         amplitudes. 
         Numbers denote effective mass in units of GeV.}
\label{fig:argand}
\end{figure}

Both the \kk{} and \ph{} elastic scattering amplitudes are plotted in 
Fig. \ref{fig:argand} in a form of Argand diagrams. 
Let us notice that a small circle of the \kk{} amplitude can be 
explained by a partial cancellation of the opposite contributions 
coming from two poles 1 and 2 related to the \am{}. 
The \ph{} results presented in Fig. \ref{fig:argand} can be compared 
with the Argand plots of Weinstein and Isgur \cite{weinstein90} and 
T\"ornqvist \cite{tornqvist95}. 
Qualitatively a shape of the \ph{} elastic amplitude is in agreement 
with Fig. 9a of \cite{weinstein90} although the inelasticity parameter 
is larger in our \ph{} amplitude above the \kk{} threshold. 
Also at higher effective \ph{} masses one sees differences which may be 
partially explained by another mass of the second \az{} meson, taken in 
Ref. \cite{weinstein90}, namely 1300 MeV, in comparison with our 
present value 1474 MeV. 
The \ph{} amplitude presented by T\"ornqvist seems to be quite similar 
to our amplitude up to the \kk{} threshold but above it one can notice 
a different behaviour. 
The most characteristic difference is a lack of a circle related to the 
\ad{} on his Argand plot (Fig. 6a in Ref. \cite{tornqvist95}). 

The \kk{} elastic amplitude plotted in Fig. 9a of Ref. 
\cite{weinstein90} is also not similar to our \kk{} amplitude. 
The inelasticity parameter $\eta$ in Fig. \ref{fig:argand} at the 
position of the \ad{} resonance is close to zero. 
This is not a case at the end point above 1400 MeV on the curve 
presented in Ref. \cite{weinstein90}. 
Briefly, the coupling of the \am{} to \kk{} is stronger in the model of 
Weinstein and Isgur than in our model. 
The opposite relation exists for the \ad{} meson. 

We have performed calculations of the production amplitudes which are 
inversely proportional to the Jost function. 
Results for the \ph{} production amplitudes are in qualitative 
agreement with those plotted in Figs. 4b and 8b by Bugg, Anisovich, Sarantsev 
and Zou for the reaction 
$p\overline{p} \to \eta \pi \pi$ \cite{bugg94}. 
\subsection{Coupling constants}

In Table \ref{tab:coupling} we give values of the coupling constants 
calculated at four poles 1 to 4 defined as in Eq. (34) of 
\cite{kaminski99}.
The ratio $|g_2^2/g_1^2|$ of the $a_0(980)$  coupling constants  
is equal to $0.72$ at the pole 1 and $0.80$ at the pole 2. Both numbers
are smaller than the value $r=1.03$ found by the Crystal Barrel Collaboration. 
Thus the width difference $\Gamma_2-\Gamma_1=18$ MeV (seen in 
Table \ref{tab:poles}) is smaller than the value $\Gamma_2-\Gamma_1=43$ 
MeV following from Eq. \eqref{cb:our_poles}. 
However, the production amplitudes calculated within our model are in 
good agreement with the amplitudes obtained in [6] using the 
modified Flatt\'e formula in the invariant mass range up to $1.05$ GeV. 
Moduli of the coupling constants corresponding to the \ad{} poles are 
stronger than the \am{} coupling constants which can explain 
a considerably larger width of the \ad{} than the width of the \am{}. 

\begin{table}[ht]
\caption{Moduli of the coupling constants $\Big|\cfrac{g_i^2}{4 \pi}\Big|$ 
         in $\rm GeV^2$}
\vspace{6pt}
\centering
\begin{tabular}{|c|c|c|c|c|} \hline
        & \multicolumn{2}{|c|}{\am{}} 
        & \multicolumn{2}{c|}{\ad{}}    \\ \cline{2-5}
channel & 1     & 2     & 3     & 4     \\ \hline
\ph{}   & 0.356 & 0.360 & 0.865 & 0.887 \\
\kk{}   & 0.256 & 0.287 & 1.078 & 1.139 \\ \hline
\end{tabular} \label{tab:coupling}
\end{table}

Despite of the closed values of the coupling constants at the poles 
1 and 2, corresponding to the resonance \am{}, one should not 
average their values since these two poles are not symmetrically 
located, with respect to the origin, in the complex plane of the 
\kk{} momentum $k_2$ (Fig. \ref{fig:poles}). 
The same remark is true for the poles 3 and 4 corresponding to the 
\ad{}. 
\subsection{Decoupled channels}
Let us now consider a limit of the uncoupled \ph{} and \kk{} channels 
in which the interchannel coupling constant $\lambda_{12}$ goes to 
zero. 
\begin{figure}[!ht]
\begin{center}
\includegraphics*[width=0.8\textwidth]{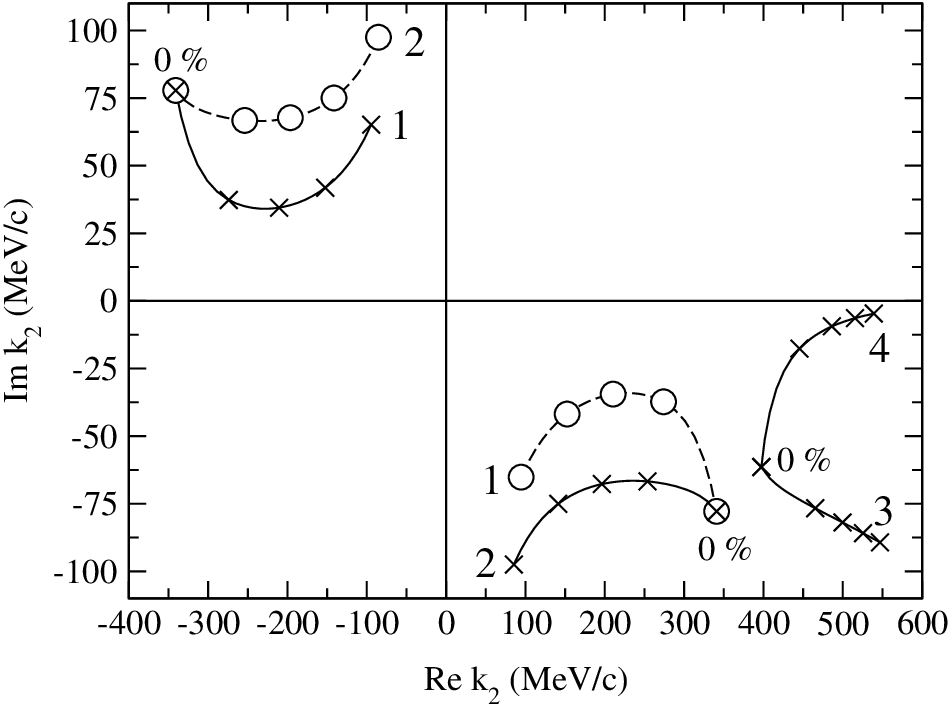}
\end{center}
\caption{Trajectories of four S--matrix poles in \kk{} complex 
         momentum plane (solid lines). 
	 Dashed lines correspond to trajectories of two zeroes of the 
	 $S_{22}$ matrix element related to \am{}. 
	 Crosses and circles on trajectories indicate points of a 
	 reduction of the interchannel coupling $\lambda_{12}^2$ by 
	 $25 \%$ steps down to $\lambda_{12}^2=0$ ($0 \%$).}
\label{fig:traj}
\end{figure}
We can gradually diminish $\lambda_{12}$ from its fitted value to a 
zero and observe trajectories of four poles (see Fig. \ref{fig:traj}). 
One can notice that in the limit $\lambda_{12} \to 0$ two poles 
3 and 4 meet together in the complex plane of $k_2$ momentum which 
means that in the uncoupled \kk{} channel 
there exist a \kk{} resonance at the effective mass equal to 
($1270.1-76.9i$) MeV. 
This \kk{} resonance evolves into the \ad{} when the coupling between 
channels is switched on. 

Behaviour of the \am{} poles is completely different. 
The pole 1 corresponding to the \am{} on sheet \pmp{} meets the 
trajectory end of the $S_{22}$ zero related to the secondary pole 2 %
\footnote{Zeroes of the $S_{22}$ matrix element at $(k_1,k_2)$ are 
related to the poles of $S_{22}$ at $(k_1,-k_2)$ and vice versa 
(see Eqs. \eqref{eq:S}).}. 
Below the real axis the pole 2 moves to a point where it meets the 
$S_{22}$ zero related to the pole 1. 
Thus effectively both poles 1 and 2 disappear from the \kk{} channel 
in the limit of the vanishing interchannel coupling constant. 
In the \ph{} channel the poles 1 and 2 meet together in this limit at 
one point, so for the parameter set \eqref{eq:par} the \am{} state 
has its origin in the \ph{} channel 
as a resonance at the effective mass ($1199.7-88.5i$) MeV. 

In such a way one can show that although the \kk{} forces in the $I=1$ 
S--wave are attractive ($\lambda_{22}<0$ in \eqref{eq:par}) they 
are not sufficiently strong to form a bound \kk{} state. 
This information about the nature of the \am{} meson can be confronted 
with the results obtained in literature on two close scalars \fm{} 
and \am{}. 
In \cite{kaminski99} one can find different solutions for the \pipi{}, 
\kk{} and $4 \pi$ amplitudes fitted to the same experimental data. 
However, the data near the \kk{} threshold are not 
yet precise enough to state whether the attractive \kk{} forces in the 
isoscalar S--wave can always create a bound \fm{} state%
\footnote{However, the \kk{} coupling constants of the \fm{} are much 
larger than the \kk{} constants of the \am{}. 
They can attain values close to $2 \;\; \rm GeV^2$ in comparison with 
values below $0.3 \;\; \rm GeV^2$ shown in Table \ref{tab:coupling}.}.  
This result is at variance with a statement of Weinstein and Isgur 
\cite{weinstein90} that both \fm{} and \am{} are the bound \kk{} 
states. 
We agree, however, with the conclusion of Janssen, Pearce, Holinde and 
Speth \cite{janssen95} that the \am{} meson is not a bound \kk{} state. 

We have studied to what extend the conclusion about 
nonexistence of the \kk{} bound state in the isovector S--wave 
remains valid in the limit of vanishing coupling constant between \ph{} 
and \kk{} channels while simultaneously the input parameters like the 
masses and the widths of the \am{} and \ad{} are modified. 
We have checked that this is true if the \am{} mass on sheet II changes between 
950 MeV and 1020 MeV, its width varies between 25 and 100 MeV, the \ad{} mass 
on sheet III changes between 1300 MeV and 1550 MeV and its width varies 
between 150 MeV and 350 MeV. 
In order to check possible large systematic errors of the input and to 
verify that the conclusion is firm the above limits are chosen larger 
than the errors of the \az{} resonance parameters given by the Particle 
Data Group. 
\section{Conclusions}

In summary, we have constructed the coupled channel model of two \az{} 
resonances decaying into the \ph{} and \kk{} mesons. 
The parameters of the separable potentials have been fitted to 
experimental values of the \az{} masses and widths and to the 
\kk{}/\ph{} branching ratio near the \kk{} threshold. 
Then we have predicted the \ad{} branching ratio and verified that it 
was in agreement with the result of the Crystal Barrel Collaboration. 
The discrepancy in the interpretation of the \am{} resonance mass 
position between the Crystal Barrel Collaboration and the E852 Group 
has been explained. 
The \am{} resonance can be described by two poles near the \kk{} 
threshold lying on sheets II and III. 
These two poles have substantially different widths when the production 
amplitudes are interpreted using the Flatt\'e model with 
non vanishing \kk{} coupling constant. 
In our model with parameters fixed by the present experimental data the 
\am{} state cannot be interpreted as a bound \kk{} state. 

\def\cb {(Crystal Barrel Coll.)}
\def\ob {(OBELIX Coll.)}
\def\wa {(WA102 Coll.)}
\def\gams {(GAMS Coll.)}
\def\bnl {(E852 Coll.)}
\def\kloe {(KLOE Coll.)}
\def\babar {(BABAR Coll.)}
\def\snd {(SND Coll.)}
\def\cmd {(CMD-2 Coll.)}
\def\pdg {(Particle Data Group)}

\end{document}